\documentclass[ amsmath, amssymb, aps, prl, twocolumn, lengthcheck, sort&compress, showpacs ]{revtex4-1}    
\usepackage{graphicx}

\usepackage{amsmath}
\usepackage{amssymb}
\usepackage{graphicx}
\usepackage{dsfont}
\usepackage{dcolumn}
\usepackage{units}
\usepackage{bm}
\usepackage{wasysym}
\usepackage{multirow}
\usepackage{slashed}
\usepackage[usenames]{color}
\usepackage[colorlinks=true,linkcolor=blue,citecolor=blue,urlcolor=blue]{hyperref}

      \newcommand{\mS}{\mathcal{S}}
      \newcommand{\mD}{\mathcal{D}}
      \newcommand{\mT}{\mathcal{T}}

      \newcommand{\conjg}[1]{\ensuremath{\hspace{1pt}\overline{\hspace{-1pt}#1\hspace{-1pt}}}\hspace{1pt}}

      \newcommand{\m}[1]{\mathbf{#1}}
      \newcommand{\cm}[1]{\overline{\mathbf{#1}}}

\begin{document}

\title{The light scalar mesons as tetraquarks}

\author{Gernot Eichmann, Christian S. Fischer, and Walter Heupel}

\affiliation{Institut f\"{u}r Theoretische Physik, Justus-Liebig-Universit\"at Giessen, 35392 Giessen, Germany}

\date{\today}

  \begin{abstract}
  We present a numerical solution of the four-quark Bethe-Salpeter equation for ground-state scalar tetraquarks
  with $J^{PC}=0^{++}$. We find that the four-body equation dynamically generates pseudoscalar-meson poles in
  the Bethe-Salpeter amplitude. The resulting tetraquarks are genuine four-quark states that are dominated
  by pseudoscalar meson-meson correlations. Diquark-antidiquark contributions are subleading because of their larger
  mass scale. In the light quark sector, the sensitivity of the tetraquark wave function to the pion poles leads
  to an isoscalar tetraquark mass $M_\sigma \sim 350$~MeV which is comparable to that of the $\sigma/f_0(500)$.
  The masses of its multiplet partners $\kappa$ and $a_0/f_0$ follow a similar pattern. This provides support
  for a tetraquark interpretation of the light scalar meson nonet in terms of 'meson molecules'.
  \end{abstract}

  \pacs{14.40.Be, 14.40.Rt, 11.10.St, 12.38.Lg}

\maketitle

  \textbf{Introduction.} ---
  The nature of the light scalar mesons $\sigma/f_0(500)$, $\kappa(800)$ and $a_0,f_0(980)$ has been an ongoing conundrum for several decades.
  Although the $\sigma$ is a central ingredient to hadronic interaction models, even its existence was experimentally not firmly established until recently.
  Taking into account evidence from new data for heavy-meson~\cite{Ablikim:2004qna} and kaon decays~\cite{Batley:2010zza},
  together with dispersive approaches to $\pi\pi$
  scattering using Roy equations~\cite{Caprini:2005zr},
  the $\sigma$ is now again listed in the PDG with a $T-$matrix pole position $(400 \dots 550) - i(200\dots 350)$ MeV~\cite{PDG}.

  Nevertheless, the $\sigma$ and its multiplet partners still do not fit well into the light $q\conjg{q}$ meson spectrum.
  In the non-relativistic quark-model classification the pseudoscalar and vector mesons ($J^{PC}=0^{-+}, 1^{--}$)
  are $s$ waves without orbital angular momentum, whereas the $0^{++}$, $1^{+-}$, $1^{++}$, and $2^{++}$
  states are $p$ waves and therefore their masses
  should be considerably larger, in contrast to
  what is found experimentally for the scalars.
  Even more puzzling is the mass degeneracy of $a_0$ and $f_0$ inside the multiplet. If they were ideally mixed $q\conjg{q}$ states,
  the iso\-singlet $\sigma$ and isotriplet $a_0$ would be mass-degenerate and consist of light quarks only, whereas $\kappa$
  would contain one strange quark and $f_0$ two.
  A $q\conjg{q}$ description would then require an enormous amount of flavor mixing to generate the observed splitting between $\sigma$ and $a_0$ and, in addition,
  to establish an accidental mass degeneracy between $a_0$ and $f_0$.
  Connected with this are the different decay properties: $a_0$ and $f_0$ are relatively narrow states close to the $K\conjg{K}$ threshold,
  whereas $\sigma$ and $\kappa$ are broad resonances that decay into $\pi\pi$ and $\pi K$, respectively.

  A natural explanation for these peculiarities was proposed long ago~\cite{Jaffe}: what if the light scalar mesons were
  tetraquarks in the form of diquark-antidiquark \mbox{($dq$-$\conjg{dq}$)} states?
  A scalar diquark forms a color-$SU(3)_c$ and flavor-$SU(3)_f$ antitriplet, and
  the combination $\cm{3}\otimes\m{3} = \m{1}\oplus\m{8}$ provides a color singlet
  and a flavor nonet.
  However, the mass ordering is reversed: the $\sigma$ would be the lightest state made of $u/d$ quarks only,
  whereas $f_0$ and $a_0$ would be heaviest and mass-degenerate because they carry two strange quarks.
  This would also explain the decay widths: $f_0$ and $a_0$ are close to $K\bar{K}$ threshold and therefore narrow, but
  $\sigma$ and $\kappa$ can simply fall apart through the gluonless OZI-superallowed decay.
  In such a scenario the true scalar $q\bar{q}$ ground states could be identified with the
  experimental `first excited nonet' with masses in the $1.3 \dots 1.5$~GeV region,
  similar to the axialvector and tensor mesons and in agreement with the nonelativistic
  estimate. The non-$q\bar{q}$ interpretation of the light scalar nonet is supported by a variety of theoretical approaches such
  as QCD sum rules~\cite{Chen:2007xr}, unitarized ChPT~\cite{Pelaez:2004xp,RuizdeElvira:2010cs},
  relativistic quark models~\cite{Ebert:2008id}, effective Lagrangians \cite{Hooft:2008we}, or the extended linear $\sigma$
  model~\cite{Parganlija:2010fz}.

  Of course, the four quarks may arrange themselves also differently: apart from the
  $dq$-$\conjg{dq}$ configuration, they could form meson molecules with two color-singlet
  clusters. In the heavy quark region, tetraquark candidates in the $XYZ$ spectrum such as the
  $X(3872)$ have been suggested to be of such molecular nature~\cite{Swanson:2003tb,XYZ}.
  In the light meson sector, the decay patterns and
  the proximity of $a_0$ and $f_0$ to the $K\conjg{K}$ threshold may also point in this
  direction~\cite{Weinstein:1990gu,molecule}: the $K\bar{K}$ molecule picture naturally explains the
  observed narrowness of these states. A potential drawback of the molecule picture for
  the light meson sector is that in potential models for light scalar molecules
  only some states are found to be bound, so that no complete nonet exists \cite{Weinstein:1990gu}.

  In this work we reconsider the question of the tetraquark nature of scalar mesons. We solve
  the relativistic four-body Bethe-Salpeter equation (BSE) for two quarks and two antiquarks with the
  overall quantum numbers of a scalar. Compared to previous approaches to the four-body system
  we improve on the following aspects:
  (a) An inherent limitation of many \mbox{$dq$-$\conjg{dq}$} and molecular models is that they assume specific
  quark configurations inside the tetraquark to make the model tractable. Depending on the
  basic assumptions on the dominating part of the underlying QCD forces these are organized
  either in terms of \mbox{$dq$-$\conjg{dq}$} or molecular flavor states, and corresponding assumptions
  on the spatial part of the wave function are made. We overcome such limitations by solving
  the four-quark BSE without any prejudice on the flavor and spatial structure of the wave function.
  (b) The framework is quantum field theoretical in nature and fully relativistic in contrast to potential models.
  (c) The \mbox{(anti-)}quarks acquire a dynamical, momentum-dependent mass via their nonperturbative
  interactions as described by the Dyson-Schwinger equation of the quark propagator. The quark-(anti-)quark
  interaction in our approach is given by nonperturbative gluon exchange in an approximation that
  satisfies the axial Ward-Takashi identity. As an important consequence, the Goldstone nature of
  the pseudoscalar mesons is preserved.

  It turns out that especially the last property is crucial for a successful description of
  the light scalar meson spectrum. Driven by the underlying quark-gluon interaction,
  meson and diquark pole structures are generated dynamically by the two-body interactions in the four-body BSE.
  Due to the color algebra, the interaction in the $q\conjg{q}$ channel is by a factor
  of two stronger than the one in the $qq$ channel. Consequently, diquarks have a larger mass scale
  than corresponding mesons with the same flavor content. In the light quark sector, dynamical chiral
  symmetry breaking greatly enhances this difference. As will be explained in detail below, these light
  meson poles dominate the resulting tetraquark wave function whereas the diquark singularities only play a very
  minor role. Thus a 'molecular picture' of light scalar mesons naturally arises. We obtain a complete
  multiplet of states including the $f_0(980)$ and $a_0(980)$ with strong $K\bar{K}$-components, but
  also the $f_0(500)$ with a dominating $\pi\pi$ component which naturally explains its large decay width.

  While we arrived at part of these conclusions already in the two-body framework of Ref.~\cite{Heupel:2012ua},
  the current work is a substantial improvement since it does not rely on many of the assumptions made
  in \cite{Heupel:2012ua}. Still, there are approximations involved to make the extremely complicated
  equations tractable. These are explained and discussed below in the technical part of the paper.
  We will argue that all approximations affect the quantitative but not the qualitative results
  of this work. In particular we believe that the molecular interpretation of the states is a robust
  feature of the framework.

  In order to avoid confusion there is an important caveat to make: Here and in the following we use the
  term 'molecule' or 'molecular picture' in a probably different sense than what is traditional in the
  literature. Working in momentum space, we do not make any assumptions on the spatial distribution
  of the (anti-)quarks inside the tetraquark. Thus we cannot, and in fact do not aim to, make a
  distinction between a state with spatially separated (anti-)diquark or meson clusters. What distinguishes
  a  `diquark-antidiquark' from a `meson molecule' picture in our framework is solely the influence of the
  corresponding analytic structure in the internal bound-state kinematics on the tetraquark wave function
  (as detailed below).

  While we exemplify our approach with the important case of scalar tetraquarks, the framework is general
  and has potential applications also in many other branches of theoretical physics. Obvious examples are
  four-body states on the nuclear or the atomic level with interesting (non-relativistic)
  applications pointed out recently \cite{Blume}.

  This work is organized as follows: In the next two sections we outline the theoretical framework and detail
  the approximations that we introduce in order to make the four-body equation tractable. We then
  present our results and discuss some implications in the conclusions.

            \begin{figure}[t!]
            \centerline{%
            \includegraphics[width=8.8cm]{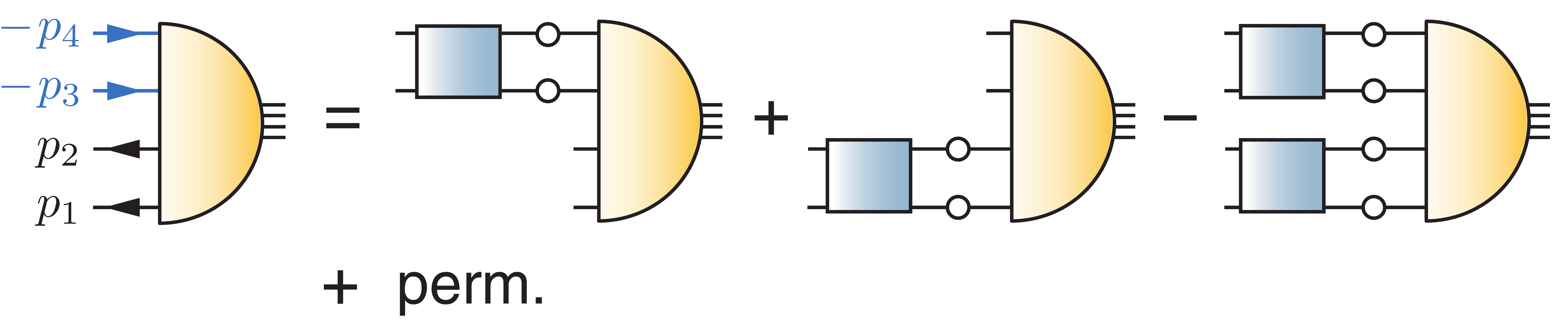}}
            \caption{Four-body BSE for the tetraquark BS amplitude. Only two-body interactions are retained, whereas irreducible three- and four-body interactions are neglected.}
            \label{fig:4b-bse}
            \end{figure}

  \smallskip

  \textbf{Bethe-Salpeter equation.} ---
  If four-quark states exist in QCD, they will appear as poles in the $qq\overline{qq}$ scattering matrix.
  The $T$ matrix satisfies an exact scattering equation
  \begin{align}
    T=K+K G_0\,T \,,
  \end{align}
  where $K$ is the four-quark interaction kernel and $G_0$ is the product of four dressed \mbox{(anti-)}quark propagators.
  The pole residue of the scattering equation is the homogeneous BSE for the four-quark Bethe-Salpeter (BS) amplitude shown in Fig.~\ref{fig:4b-bse}.
  In compact notation it is written as
  \begin{align}
   \Gamma = K G_0\,\Gamma  \,,
  \end{align}
  which has a solution only if the $T$ matrix has a pole.
  The exact kernel $K$ is the sum of two-, three- and four-body irreducible interactions.
  In Fig.~\ref{fig:4b-bse} we already use an approximated kernel, where we omitted all irreducible
  three- and four-body interactions (the resulting equation can be rewritten as a Faddeev-Yakubovsky
  equation~\cite{faddeev-yakubowski}).

  This approximation is severe and can be disputed. Certainly, in a more complete approach, these terms need
  to be analyzed in detail and taken into account accordingly. The main reason to omit these terms here
  is simplicity. As will become clear below, solving the relativistic four-body equation is a tremendous
  numerical task and some simplifications have to be made in order to make the equations tractable. However,
  we also believe that these approximations receive an important justification a posteriori. As already stated
  in the introduction, we find that the structure of the tetraquark wave functions are strongly dominated
  by the meson poles in the two-body interaction channels. As a consequence, the masses of the resulting bound
  states rely almost completely on the location of these poles. This feature of the four-body BSE will not be
  changed by three-body or four-body interactions, unless these introduce new analytic structures into the equation
  that have a stronger influence on the integration region of the BSE than the meson poles. No such structures
  are known in the three-\mbox{(anti)}quark and four-\mbox{(anti)}quark channels. We therefore believe that the emerging `meson molecule'
  picture, in the sense explained in the introduction, is robust with respect to this approximation.

  %
  In addition, we note that the introduction of irreducible four-quark interactions a priori does not have any
  prejudice towards a \mbox{$dq$-$\conjg{dq}$} or meson molecule picture. For example, the instanton-induced six-fermion
  effective interaction used in Ref.~\cite{Hooft:2008we} introduces a tetraquark-$q\bar{q}$ mixing term that
  has (anti-)diquark channels for the tetraquark states but also terms where the tetraquark is in its meson-meson
  flavor state. Pending detailed investigations, this must be a general feature of the four-body kernel.

  Coming back to Fig.~\ref{fig:4b-bse}, the specific form of the remaining two-body interactions is necessary to prevent overcounting~\cite{Huang,Khvedelidze:1991qb,Heupel:2012ua}.
  The equation in the figure is written in the \mbox{$dq$-$\conjg{dq}$} topology (12)(34);
  there are two further permutations (23)(14) and (31)(24)
  with meson-meson topologies.
  The two-body kernels that appear in Fig.~\ref{fig:4b-bse} must be consistent with the underlying quark-gluon structure to preserve QCD's chiral symmetry.
  This is achieved by employing a rainbow-ladder kernel,
  where the $qq$ and $q\conjg{q}$ interaction is generated
  by iterated non-perturbative gluon exchange. The Dyson-Schwinger
  equation for the quark propagator is solved with this interaction.
  All relevant formulas and input values are given in
  Refs.~\cite{Eichmann:2009qa,Eichmann:2011vu} where the approach is applied to baryons; we will not repeat them here for brevity.
  The rainbow-ladder setup describes the phenomenology of pseudoscalar and vector mesons as well as baryon octet and decuplet ground states
  reasonably well; see~\cite{RL} for reviews.
  This implies not only mass spectra but also their form factors and other properties,
  and it extends to charmonium and bottomonium spectra~\cite{Fischer:2014cfa}.

  One should note that rainbow-ladder alone does \textit{not} provide satisfactory results for
  scalar and axialvector mesons (the `$p$ waves' in the quark model).
  Viewed as a $q\bar{q}$ state, rainbow-ladder produces a $\sigma-$meson mass of $600 \dots 700$~MeV~\cite{Alkofer:2002bp},
  with potentially sizeable corrections beyond rainbow-ladder~\cite{BRL-sigma} contrary to the masses of pseudoscalar and vector mesons.
  In any case, since these calculations support ideal flavor mixing they still lead to the mass ordering $\{\sigma, a_0\} \rightarrow \kappa \rightarrow f_0$,
  which means there must be further effects at play.
  On the other hand, a scalar $qq\conjg{q}\conjg{q}$ state \textit{does} form an $s-$wave orbital ground state.
  Hence, employing a rainbow-ladder kernel in the four-body equation (as opposed to a scalar $q\conjg{q}$ meson) is well motivated from the aforementioned meson and baryon studies and its reliability can be judged from their results.

  \smallskip

  \textbf{Tetraquark amplitude.} ---
  The $0^{+}$ tetraquark BS amplitude has the form $\Gamma_{\alpha\beta\gamma\delta}(p,q,k,P)$,
  where Greek subscripts are Dirac indices, $P$ is the total momentum with $P^2=-M^2$, and $M$ is the tetraquark mass.
  A convenient choice for the three relative momenta
  are the Mandelstam momenta in the meson-meson and \mbox{$dq$-$\conjg{dq}$} configurations, i.e., the respective sums  of the pair momenta:
  \begin{equation} \renewcommand{\arraystretch}{1.3}
  \begin{array}{rl}
      p &= \tfrac{1}{2}\,(p_2+p_3-p_1-p_4)\,, \\
      q &= \tfrac{1}{2}\,(p_3+p_1-p_2-p_4)\,, \\
      k &= \tfrac{1}{2}\,(p_1+p_2-p_3-p_4)\,,
  \end{array}\quad P=\sum_{i=1}^4 p_i\,.
  \end{equation}
  Apart from color and flavor, the Dirac part of the full amplitude is constructed from 256 tensor structures:
            \begin{equation}
                \Gamma(p,q,k,P) = \sum_{n=1}^{256} f_{n}(\dots)\,\tau_{n}(p,q,k,P) \,.
            \end{equation}
  The BSE leads to a system of coupled integral equations for the dressing functions $f_n$ which
  depend on nine Lorentz invariants: the `Mandelstam variables' $p^2$, $q^2$, $k^2$
  and six further angular variables ($p\cdot q$, $p\cdot P$, etc.).
  This is also the main difficulty in solving the BSE numerically:
  even with only 10 points for each Lorentz invariant, each dressing function would be defined on $10^{9}$ grid points.

\begin{figure}[t!]
       \centerline{%
       \includegraphics[width=7cm]{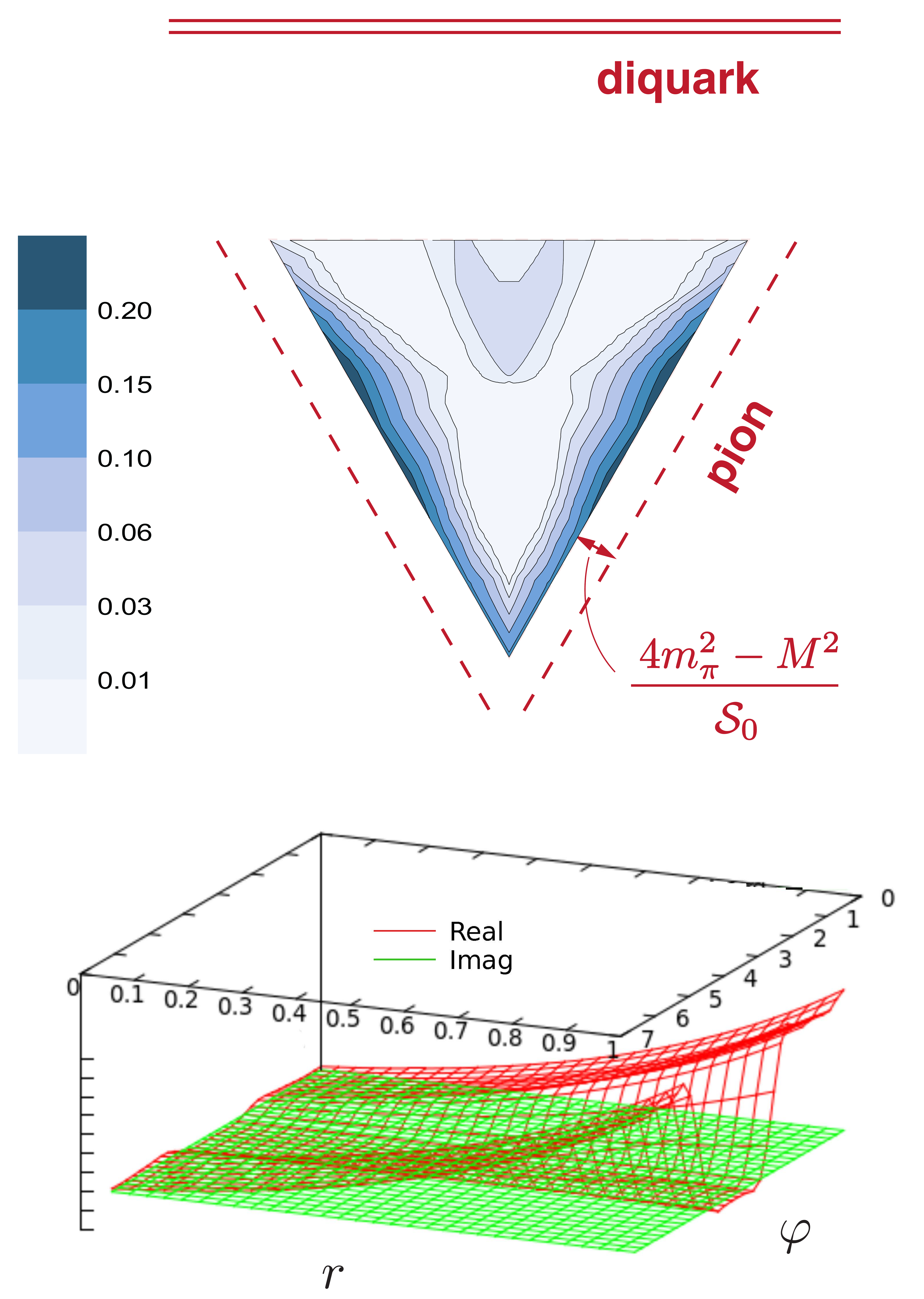}}
       \caption{\textit{Top:} Mandelstam triangle whose axes (not shown) are the doublet variables in Eq.~\eqref{multiplets}.
                The top/right/left edge of the triangle corresponds to vanishing $k^2$, $p^2$ or $q^2$, respectively.
                The contour plot is representative and shows the magnitude of the leading dressing function. \textit{Bottom:} Same plot in polar coordinates, where the radius $0 < r < 1$
                measures the distance from the center to the edges. The gap is due to missing diquark poles. }
       \label{fig:doublet}
       \end{figure}

  To make the problem tractable, we arrange the Lorentz invariants into multiplets of the permutation group $S_4$~\cite{Eichmann:2015nra}.
  This allows us to switch on groups of variables separately in the solution process
  and thereby judge their importance without destroying the symmetries of the system.
  The Mandelstam variables can be grouped into a symmetric singlet $\mathcal{S}_0$ and a doublet $\mD$,
  \begin{equation}\label{multiplets}
      \mS_0 = \frac{p^2+q^2+k^2}{4}\,, \quad \mD = \frac{1}{4\mS_0}\left[\begin{array}{c} \sqrt{3}\,(q^2-p^2) \\ p^2+q^2-2k^2 \end{array}\right],
  \end{equation}
  whereas the remaining six angular variables form two triplets $\mT_1$, $\mT_2$.
  Hence, the dressing functions can be written as $f_n=f_n(\mS_0, \mD, \mT_1, \mT_2)$.
  The doublet phase space that remains invariant under the equation forms the interior of a triangle bound by the lines $p^2=0$, $q^2=0$ and $k^2=0$ (see Fig.~\ref{fig:doublet}),
  whereas the triplets form a tetrahedron and a sphere, cf.~Ref.~\cite{Eichmann:2015nra}.

   To construct the Dirac tensor basis, we orthogonalize the momenta $p,q,k,P$ to obtain
   four orthonormal momenta $n_i^\mu$ ($i=1\dots 4$) that are mutually transverse. The set
   \begin{equation}\label{basis-set-1}
       \{ \mathds{1}\,, \;\; \slashed{n}_i\,, \;\; \slashed{n}_i \,\slashed{n}_j\,, \;\; \slashed{n}_i\,\slashed{n}_j\,\slashed{n}_k\,, \;\; \slashed{n}_i\,\slashed{n}_j\,\slashed{n}_k\,\slashed{n}_l \}
   \end{equation}
   with $i<j<k<l$ consists of 16 elements; commutators are not necessary because $\slashed{n}_i\,\slashed{n}_j = -\slashed{n}_j\,\slashed{n}_i$.
   Taking all tensor products of Eq.~\eqref{basis-set-1} with itself yields $256$ independent tensor structures.
   On the other hand,
   with the help of
   \begin{equation}
       \slashed{n}_i\,\slashed{n}_j\,\slashed{n}_k\,\slashed{n}_l = -\varepsilon \gamma_5\,, \quad \varepsilon := \varepsilon^{\mu\nu\rho\sigma}\,n_i^\mu\,n_j^\nu\,n_k^\rho\,n_l^\sigma\,,
   \end{equation}
   where $\{ijkl\}$ is an even permutation of $\{1234\}$, all elements with three or four slashes can be reduced to those with two at most,
   which leaves eight elements in Eq.~\eqref{basis-set-1}.
   If we write  $\Omega_1=\mathds{1}$ and $\Omega_2=\varepsilon\gamma_5$, identify $n_4^\mu = \hat{P}^\mu$ with the normalized total momentum, and
   express $\slashed{n}_4$ in terms of the positive/negative-energy projectors $\Lambda_\pm =(\mathds{1} \pm \slashed{n}_4)/2$,
   then a complete, orthonormal, 256-dimensional positive-parity basis for the BS amplitude
   is given by
   \begin{equation}\label{basis-full-1}
       \tau_n(p,q,k,P) = \Gamma_i\,\Lambda_\lambda \,\Omega_\omega \,\gamma_5 C\otimes C^T\gamma_5\,\Omega_{\omega'}\,\Lambda_{\lambda'}\,\Gamma_j
   \end{equation}
   with $\Gamma_i \in \{ \mathds{1}, \, \slashed{n}_1, \slashed{n}_2, \slashed{n}_3 \}$.
   We inserted the combination $\gamma_5 C \otimes C^T\gamma_5$ for the \mbox{$dq$-$\conjg{dq}$} topology (12)(34);
   all further structures such as $\gamma^\mu C\otimes C^T\gamma^\mu$ but also those in the meson-meson topologies are linearly dependent.

       In practice we are interested in partial-wave bases whose elements are eigenstates
       of the total quark spin and orbital angular momentum in the tetraquark's rest frame.
       The construction is analogous to the partial-wave analysis of the nucleon's Faddeev amplitude~\cite{Eichmann:2011vu}.
       Since the tetraquark carries total angular momentum $J=0$, the spin eigenvalue coincides with the orbital angular momentum ($s=l$).
       The analysis yields $32$ $s$ waves, $144$ $p$ waves and $80$ $d$ waves.
       To obtain the $s$ waves, replace $\Gamma_i \otimes \Gamma_j$ in Eq.~\eqref{basis-full-1} by
       $\mathds{1}\otimes\mathds{1}$ and $\tfrac{1}{\sqrt{3}}\,\gamma_T^\mu \otimes \gamma_T^\mu$, where $\gamma^\mu_T = \gamma^\mu - n_4^\mu\,\slashed{n}_4$
       is transverse to the total momentum.
       This gives $2\times 16=32$ tensor structures.
       For the calculation we keep the 16 dominant $s$~waves that do not depend on any relative momentum, i.e., those with $\Omega_\omega=\Omega_{\omega'} \in \{ \mathds{1}, \gamma_5 \}$.
        This is a closed set under Fierz transformations: it contains
        \mbox{$dq$-$\conjg{dq}$} structures such as those formed by scalar and axialvector diquarks,
        \begin{equation*}
            (\gamma_5 C)_{\alpha\beta}\,(C^T\gamma_5)_{\gamma\delta}\,, \quad
            (\gamma^\mu C)_{\alpha\beta}\,(C^T\gamma^\mu)_{\gamma\delta}\,, \quad\dots
        \end{equation*}
        but by Fierz identities it automatically also includes all meson-meson structures in the crossed channels:
        \begin{equation*}
            (\gamma_5)_{\alpha\gamma}\,(\gamma_5)_{\beta\delta}\,, \quad
            (\gamma^\mu)_{\alpha\gamma}\,(\gamma^\mu)_{\beta\delta}\,, \quad \text{etc.}
        \end{equation*}
        The associated error of neglecting the $p-$ and $d-$wave tensors is discussed in the results section.

        Concerning color and flavor, the $SU(3)$ decomposition
        \begin{equation}
        \begin{split}
            & \m{3}\otimes\m{3}\otimes \cm{3} \otimes \cm{3} = ( \cm{3} \oplus \m{6}) \otimes (\m{3}\otimes\cm{6})\\
             &=\m{1} \oplus \m{1} \oplus \m{8} \oplus \m{8} \oplus \m{8} \oplus \m{8}  \oplus \m{10} \oplus \cm{10} \oplus \m{27}
        \end{split}
        \end{equation}
        provides two singlets via $\cm{3}\otimes\m{3}$ and $\m{6}\otimes\cm{6}$ (or, in the meson-meson configurations, $\m{1}\otimes\m{1}$ and $\m{8}\otimes\m{8}$).
        The complete list can be found in~\cite{Santopinto:2006my,cf} or reconstructed from the Clebsch-Gordan tables in Refs.~\cite{Kaeding:1995vq,deSwart:1963gc}; alternatively, the
        program of Ref.~\cite{Alex:2010wi} can be used.
        Even though the interaction in the color-sextet diquark channel
        is repulsive, we take into account both color singlets because they are necessary for Fierz completeness.
        Regarding flavor, all of the above multiplets may contribute in general and those with same isospin and hypercharge can mix. In our case we can simply factor out the flavor structure because
        the rainbow-ladder kernel is flavor-blind;
        different flavor states will show up as excited states in the BSE solution.

  \smallskip
  \textbf{Results.} --- In practice the BSE is solved as an eigenvalue equation, schematically written as $K G_0\,\Gamma_i = \lambda_i(P^2)\,\Gamma_i$.
        The largest eigenvalue describes the ground state and the remaining ones the excited states. Upon calculating the eigenvalue spectrum,
        $\lambda_i(P^2=-M^2)=1$ recovers the original equation and thereby allows one to extract the mass and BS amplitude of the state.

\begin{figure}[t!]
            \centerline{%
            \includegraphics[width=8.7cm]{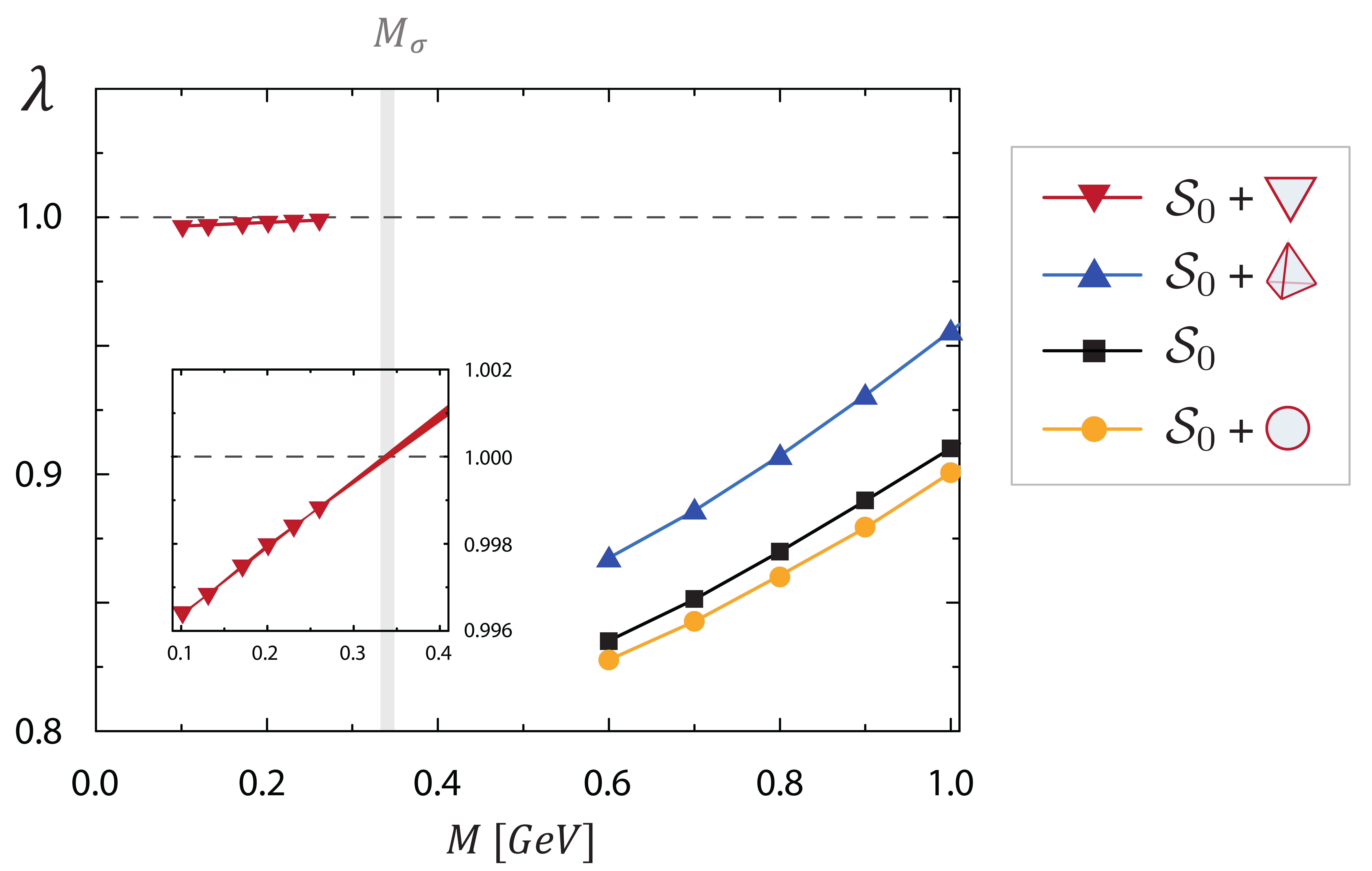}}
            \caption{Largest BSE eigenvalue for the tetraquark with four light quarks after switching on different multiplets. The band in the inset shows the result of  linear and quadratic extrapolations;
                     the resulting vertical bar gives the extracted $\sigma$ mass.}
            \label{fig:ev}
            \end{figure}

        The multiplet analysis around Eq.~(\ref{multiplets}) provides us with a convenient
        tool to isolate the relevant momentum regions probed by the equation.
        The largest eigenvalue for the system consisting of four light quarks is shown in Fig.~\ref{fig:ev}.
        The second curve from the bottom shows the result obtained by restricting the momentum dependence to $\mS_0$ only, i.e., $f_n(\mS_0, \mD, \mT_1,\mT_2) \approx f_n(\mS_0)$.
        The corresponding bound state mass is $M \sim 1.5$ GeV,
        which is essentially `four times the constituent-quark mass'.
        Taking into account either of the triplets
        does not change this behavior significantly. However, implementing the doublet has a drastic effect: the eigenvalue is now almost flat and close to 1,
        with a crossing at $\sim 350$ MeV.

        Therefore, it is the doublet phase space that carries the relevant momentum dependence. This can be understood from Fig.~\ref{fig:4b-bse}:
        upon iteration, the BSE will generate two-body $q\bar{q}$ and $qq$ scattering matrices with their own (meson and diquark) singularity structures.
        These two-body poles appear at timelike values of the Mandelstam variables, i.e., in the exterior of the doublet triangle in Fig.~\ref{fig:doublet}.
        The onset of the pion poles at $p^2<0$ and $q^2<0$ can be clearly seen in the contour plot as well as in the 3D plot of Fig.~\ref{fig:doublet}:
         the pion mass is small and the pion poles are close to the triangle, whereas the diquarks have no visible effect because their mass scales are much larger: $m_\text{sc} \sim 800$ MeV
        and $m_\text{av}\sim 1$ GeV in rainbow-ladder~\cite{fn2}.

        This feature is what drives the tetraquark mass towards low values: although the four-quark equation knows a priori nothing about mesons or diquarks,
        it generates their pole structures dynamically and these influence the behavior of the dressing functions from the exterior of the integration domain.
        Since the system is driven by the would-be pseudoscalar Goldstone bosons, the tetraquark mass can be understood as a remnant of spontaneous chiral symmetry breaking.
        On the other hand, this also naturally leads to a meson-molecule interpretation (in the sense explained in the introduction)
        simply because the diquark mass scales are too large to be relevant.

        The smallness of the pion mass also generates a physical threshold: if $M>2m_\pi$, the poles enter the integration domain and the tetraquark becomes a resonance.
        In that case the tetraquark pole in the four-quark scattering matrix will disappear from the real axis and ultimately one should solve the BSE in the complex plane,
        also taking into account the poles in the integration domain.
        This is, however, a rather formidable task which has not even been accomplished in simpler systems so far.
        Hence, for the moment we resort to simple extrapolations to estimate the real part of the mass, although
        we are aware that crossing thresholds leads to non-analyticities~\cite{Hanhart:2014ssa} and
        the results of extrapolations should be interpreted with great care.

            \begin{figure}[t!]
            \centerline{%
            \includegraphics[width=9cm]{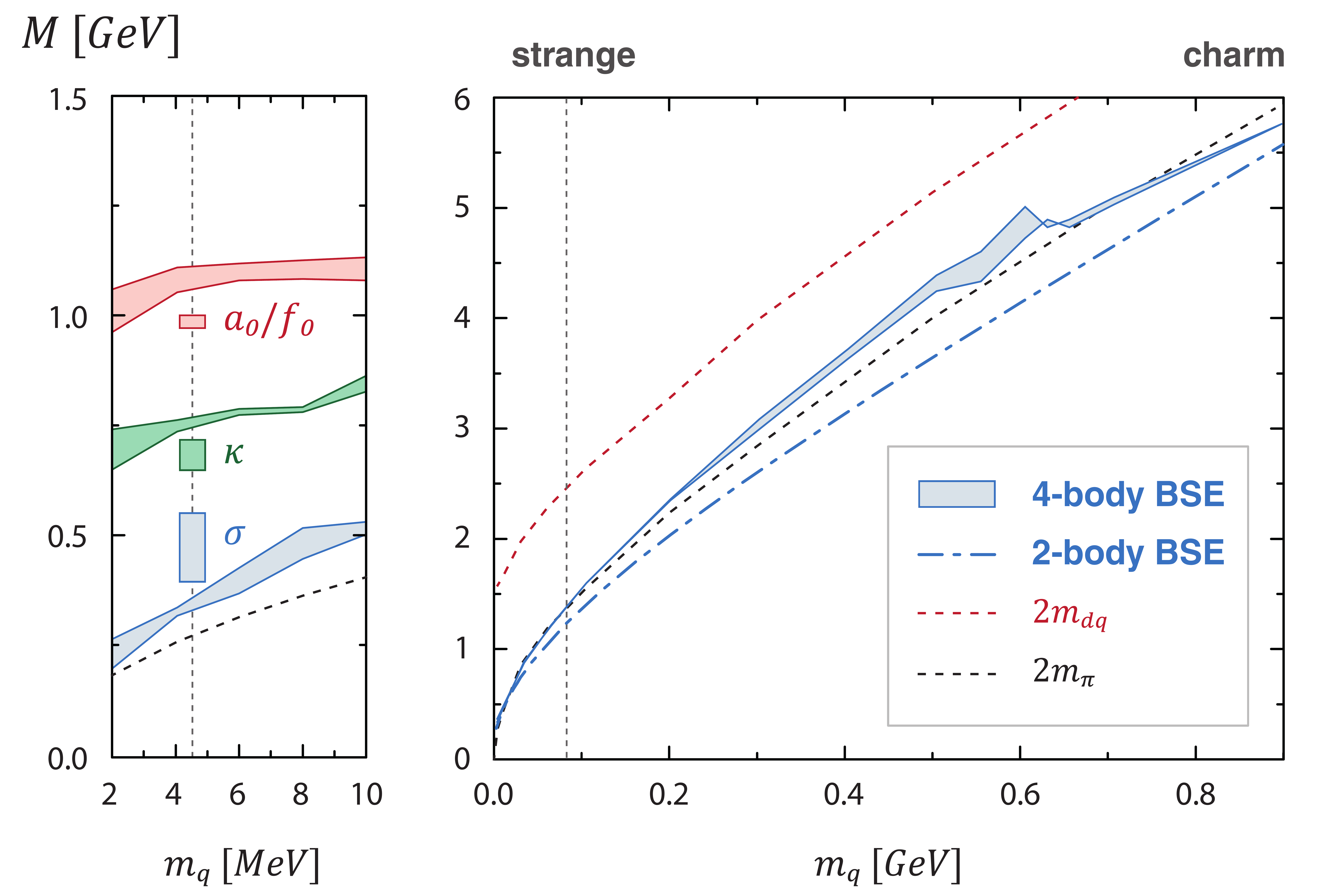}}
            \caption{\textit{Left panel:} tetraquark masses at fixed strange quark mass and varying light quark mass.
                     The bands contain the extrapolation errors and the
                     dotted curve is the two-pion threshold.
                     The vertical dotted line shows the physical $u/d$ mass and the vertical bars are the experimental values for the masses~\cite{PDG}.
                     \textit{Right panel:} quark-mass dependence of the tetraquark mass with four identical quarks.
                     The band is the present result from the four-body equation with extrapolation errors,
                     compared to that obtained with coupled meson-meson/\mbox{$dq$-$\conjg{dq}$} equations (dash-dotted curve)~\cite{Heupel:2012ua}.
                     The dotted curves are the respective thresholds.  }
            \label{fig:masses}
            \end{figure}

        To minimize numerical artifacts due to the arising pole structures,
        we approximate the dressing functions of the tetraquark BS amplitude by
        \begin{equation}
            f_n(\dots) \approx \sum_{j=1}^3 \frac{f_n^{(j)}(\mS_0,R)}{\left( q_j^2 + m_j^2 - \frac{M^2}{4}\right)^2 - (q_j\cdot P)^2}\,,
        \end{equation}
        where $q_j \in \{p,q,k\}$ are the three Mandelstam momenta, $m_j$ are the corresponding meson and diquark pole masses that appear in the doublet, and
        $R$ is the radius of the tetrahedron which forms the triplet $\mT_1$.
        The denominator is the product of two-body poles and provides an accurate representation of the momentum dependence in the Mandelstam plane (Fig.~\ref{fig:doublet});
        the pole residues are determined dynamically in the BSE solution.
        In addition to the $\mS_0$ dependence, the numerator also captures
        the subleading effects provided by the tetrahedron that are visible in Fig.~\ref{fig:ev}.

        The resulting current-quark mass dependence of the $\sigma$ mass is shown in the right panel of Fig.~\ref{fig:masses}. The tetraquark
        is a resonance just slightly above $\pi\pi$ threshold; it only becomes a bound state in the charm-quark region.
        The mass is also larger than our earlier result in Ref.~\cite{Heupel:2012ua} obtained with the coupled meson-meson/\mbox{$dq$-$\conjg{dq}$} equations (also shown in Fig.~\ref{fig:masses}).
        However, this might be due to our omission of the remaining
        $p-$ and $d-$wave tensor structures which are likely to
        provide $\sim 10\%$ effects in analogy to three-quark systems \cite{Eichmann:2009qa}.
        Analogous conclusions apply for the multiplet partners
        of the $\sigma$, shown in the left panel of Fig.~\ref{fig:masses}; the $\kappa$ is dominated by $K\pi$
        poles and $a_0/f_0$ by $\eta\pi$ and $K\conjg{K}$ poles.
        Our results at physical current-quark masses are
        \begin{equation}
           M_\sigma = 0.348(13)\,\text{GeV}
        \end{equation}
        together with
        \begin{equation}
        \begin{split}
            M_\kappa &= 0.750(12)\,\text{GeV},\\
            M_{a_0/f_0} &= 1.081(28)\,\text{GeV},
        \end{split}
        \end{equation}
        where the errors are due to the chosen extrapolation procedure (linear or quadratic).

  \smallskip
  \textbf{Summary and conclusions.} ---        To summarize, we presented a numerical solution of the four-quark Bethe-Salpeter equation for scalar tetraquarks,
        implementing a well-established rainbow-ladder (gluon-exchange) interaction.
        The resulting masses for the $\sigma$, $\kappa$ and $a_0/f_0$ are in the ballpark
        of their experimental values.
        Barring further mixing effects, this provides support
        for a tetraquark interpretation of the light scalar mesons.
        Our calculation also illustrates how a dynamical generation of resonances emerges from the quark level:
        pseudoscalar-meson poles are generated in the solution process and dominate the dynamics of the system;
        if they enter the integration domain the tetraquark becomes a resonance.
        In that sense these tetraquarks are not diquark-antidiquark states but predominantly `meson molecules',
        which explains their mass ordering and their decay widths into the hadronic channels
        as an indirect consequence of QCD's spontaneous chiral symmetry breaking.

        While we believe that our calculation makes a tetraquark interpretation of the light scalar meson
        sector further plausible, much can be done to further test and improve the framework. On systematic
        grounds it would be highly interesting to study tetraquarks with other quantum numbers. Within the
        currently employed framework and its approximations this is feasible in principle, but it requires huge
        computational resources. Furthermore, one would like to lift some of the approximations made so
        far. Expanding the framework to include also the $p-$ and $d-$wave tensor structures is conceptually
        straightforward, but, again, cost-intensive in terms of CPU time. The introduction of additional
        three-body and four-body irreducible forces requires some conceptional effort; work in this direction
        is underway.

        An important future application of our approach are tetraquarks in
        the charmonium region. One of the most interesting problems there is
        to identify the internal structure of potential candidates, i.e. to
        decipher whether they are dominated by diquark-antidiquark ($nc$-$\conjg{nc}$),
        meson molecule ($n\conjg{c}$-$\conjg{n}c$) or hadrocharmonium
        ($n\conjg{n}$-$c\conjg{c}$) configurations. These correspond to poles
        along the edges of the Mandelstam triangle in Fig.~\ref{fig:doublet}.
        Our approach has the potential to answer this question dynamically from
        the underlying quark-gluon interaction, thus bypassing the assumptions
        made by many models.

  \smallskip
  \textbf{Acknowledgments.} --
         This work was supported by the German Science Foundation DFG under
         project number DFG TR-16, by the BMBF under contract No. 06GI7121,
         by the Helmholtz International Center for FAIR within the LOEWE
         program of the State of Hesse and by the Helmholtzzentrum GSI.


\begin{thebibliography}{99}

   \small

\bibitem{Ablikim:2004qna}
  M.~Ablikim {\it et al.} [BES Collaboration],
  Phys.\ Lett.\ B {\bf 598}, 149 (2004)

\bibitem{Batley:2010zza}
  J.~R.~Batley {\it et al.} [NA48-2 Collaboration],
  Eur.\ Phys.\ J.\ C {\bf 70}, 635 (2010).


\bibitem{Caprini:2005zr}
  I.~Caprini, G.~Colangelo and H.~Leutwyler,
  Phys.\ Rev.\ Lett.\  {\bf 96}, 132001 (2006);
  R.~Garcia-Martin, R.~Kaminski, J.~R.~Pelaez and J.~Ruiz de Elvira,
  Phys.\ Rev.\ Lett.\  {\bf 107}, 072001 (2011).


\bibitem{PDG}
  K.A.~Olive \textit{et al.} [Particle Data Group], Chin.\ Phys.\ C {\bf 38}, 090001 (2014).


\bibitem{Jaffe}
  R.~L.~Jaffe,  Phys.\ Rev.\ D {\bf 15}, 267 (1977);
  F.~E.~Close and N.~A.~Tornqvist, J.\ Phys.\ G {\bf 28}, R249 (2002);
  L.~Maiani, F.~Piccinini, A.~D.~Polosa and V.~Riquer, Phys.\ Rev.\ Lett.\  {\bf 93}, 212002 (2004).


\bibitem{Weinstein:1990gu}
  J.~D.~Weinstein and N.~Isgur, Phys.\ Rev.\ D {\bf 41} (1990) 2236;

\bibitem{molecule}
  F.~E.~Close, N.~Isgur and S.~Kumano, Nucl.\ Phys.\ B {\bf 389} (1993) 513;
  T.~E.~O.~Ericson and G.~Karl, Phys.\ Lett.\ B {\bf 309}, 426 (1993);
  N.~A.~Tornqvist, Z.\ Phys.\ C {\bf 61}, 525 (1994);
  C.~Amsler and N.~A.~Tornqvist,
  Phys.\ Rept.\  {\bf 389}, 61 (2004).

\bibitem{Swanson:2003tb}
  E.~S.~Swanson,
  Phys.\ Lett.\ B {\bf 588}, 189 (2004).

\bibitem{XYZ}
   N.~Brambilla \textit{et al.}, Eur.\ Phys.\ J.\ C {\bf 74}, 2981 (2014).

\bibitem{Chen:2007xr}
  H.~X.~Chen, A.~Hosaka and S.~L.~Zhu,
  Phys.\ Rev.\ D {\bf 76}, 094025 (2007).


\bibitem{Pelaez:2004xp}
  J.~R.~Pelaez,
  Mod.\ Phys.\ Lett.\ A {\bf 19}, 2879 (2004).


\bibitem{RuizdeElvira:2010cs}
  J.~Ruiz de Elvira, J.~R.~Pelaez, M.~R.~Pennington and D.~J.~Wilson,
  Phys.\ Rev.\ D {\bf 84}, 096006 (2011).


\bibitem{Ebert:2008id}
  D.~Ebert, R.~N.~Faustov and V.~O.~Galkin,
  Eur.\ Phys.\ J.\ C {\bf 60}, 273 (2009).

\bibitem{Hooft:2008we}
G.~'t Hooft, G.~Isidori, L.~Maiani, A.~D.~Polosa and V.~Riquer,
Phys.\ Lett.\ B {\bf 662} (2008) 424


  \bibitem{Parganlija:2010fz}
  D.~Parganlija, F.~Giacosa and D.~H.~Rischke,
  Phys.\ Rev.\ D {\bf 82}, 054024 (2010).


\bibitem{Heupel:2012ua}
  W.~Heupel, G.~Eichmann and C.~S.~Fischer,
  Phys.\ Lett.\ B {\bf 718}, 545 (2012).

\bibitem{Blume}
 D.~Blume, Phys.\ Rev.\ Lett.\ {\bf 109}, 230404 (2012).

\bibitem{faddeev-yakubowski}
  O.~A.~Yakubovsky, Sov.\ J.\ Nucl.\ Phys.\  {\bf 5}, 937 (1967).


  \bibitem{Huang}
  K.~Huang and H.~A.~Weldon,
  Phys.\ Rev.\ D {\bf 11}, 257 (1975).

\bibitem{Khvedelidze:1991qb}
  A.~M.~Khvedelidze and A.~N.~Kvinikhidze,
  Theor.\ Math.\ Phys.\  {\bf 90}, 62 (1992).




  \bibitem{Eichmann:2009qa}
  G.~Eichmann, R.~Alkofer, A.~Krassnigg and D.~Nicmorus,
  Phys.\ Rev.\ Lett.\  {\bf 104}, 201601 (2010).

\bibitem{Eichmann:2011vu}
  G.~Eichmann,
  Phys.\ Rev.\ D {\bf 84}, 014014 (2011).

\bibitem{RL}
  P.~Maris and P.~C.~Tandy, Nucl.\ Phys.\ Proc.\ Suppl.\  {\bf 161}, 136 (2006);
  G.~Eichmann, Prog.\ Part.\ Nucl.\ Phys.\  {\bf 67}, 234 (2012);
  I.~C.~Cloet and C.~D.~Roberts, Prog.\ Part.\ Nucl.\ Phys.\  {\bf 77}, 1 (2014);
  H.~Sanchis-Alepuz and R.~Williams, arXiv:1503.05896 [hep-ph].




  \bibitem{Fischer:2014cfa}
  C.~S.~Fischer, S.~Kubrak and R.~Williams,
  Eur.\ Phys.\ J.\ A {\bf 51}, 10 (2015);
  T.~Hilger, C.~Popovici, M.~Gomez-Rocha and A.~Krassnigg,
  Phys.\ Rev.\ D {\bf 91}, no. 3, 034013 (2015).


\bibitem{Alkofer:2002bp}
  R.~Alkofer, P.~Watson and H.~Weigel,
  Phys.\ Rev.\ D {\bf 65}, 094026 (2002).


  \bibitem{BRL-sigma}
  R.~Williams and C.~S.~Fischer, Chin.\ Phys.\ C {\bf 34}, no. 9, 1500 (2010);
  L.~Chang, C.~D.~Roberts and P.~C.~Tandy, Chin.\ J.\ Phys.\  {\bf 49}, 955 (2011);
  H.~Sanchis-Alepuz and R.~Williams, arXiv:1504.07776 [hep-ph].

\bibitem{Eichmann:2015nra}
  G.~Eichmann, C.~S.~Fischer and W.~Heupel,
  arXiv:1505.06336 [hep-ph].

\bibitem{Santopinto:2006my}
  E.~Santopinto and G.~Galata,
  Phys.\ Rev.\ C {\bf 75}, 045206 (2007).

\bibitem{cf}
  We follow the conventions of de Swart~\cite{deSwart:1963gc}; in that case one has to replace $\{d,u,s\}$ in Ref.~\cite{Santopinto:2006my}
  by $\{u,d,s\}$ and $\{-\conjg{u}, -\conjg{d}, \conjg{s}\}$ by $\{\conjg{d}, -\conjg{u}, \conjg{s}\}$.

\bibitem{deSwart:1963gc}
  J.~J.~de Swart,
  Rev.\ Mod.\ Phys.\  {\bf 35}, 916 (1963).

\bibitem{Kaeding:1995vq}
  T.~A.~Kaeding,
  Atom.\ Data Nucl.\ Data Tabl.\  {\bf 61}, 233 (1995).

\bibitem{Alex:2010wi}
  A.~Alex, M.~Kalus, A.~Huckleberry and J.~von Delft,
  J.\ Math.\ Phys.\  {\bf 52}, 023507 (2011).

\bibitem{fn2}
  P.~Maris,
  Few Body Syst.\  {\bf 32}, 41 (2002).
  In a rainbow-ladder approximation the quark-quark scattering matrix features bound-state poles,
  but implementing crossed-ladder exchange removes them from the real axis as shown in: 
  A.~Bender, C.~D.~Roberts and L.~Von Smekal,
  Phys.\ Lett.\ B {\bf 380}, 7 (1996).

\bibitem{Hanhart:2014ssa}
  C.~Hanhart, J.~R.~Pelaez and G.~Rios,
  Phys.\ Lett.\ B {\bf 739} (2014) 375.




\end{thebibliography}
\end{document}